\documentclass[twocolumn,aps,prl,amsmath,amssymb,showpacs,preprintnumbers]{revtex4}
\usepackage{graphicx}
\begin{document}


\title{Large scale effective theory for cosmological bounces}

\author{Martin Bojowald}
\email{bojowald@gravity.psu.edu}
\affiliation{Institute for Gravitational Physics and Geometry,
The Pennsylvania State University, 104 Davey Lab, University Park,
PA 16802, USA}



\newcommand{\lP}{\ell_{\mathrm P}}
\newcommand{\vp}{\varphi}
\newcommand{\vt}{\vartheta}

\newcommand{\md}{{\mathrm{d}}}
\newcommand{\tr}{\mathop{\mathrm{tr}}}
\newcommand{\sgn}{\mathop{\mathrm{sgn}}}

\newcommand*{\R}{{\mathbb R}}
\newcommand*{\N}{{\mathbb N}}
\newcommand*{\Z}{{\mathbb Z}}

\begin{abstract}
  An exactly solvable bounce model in loop quantum cosmology is
  identified which serves as a perturbative basis for realistic bounce
  scenarios. Its bouncing solutions are derived analytically,
  demonstrating why recent numerical simulations robustly led to
  smooth bounces under the assumption of semiclassicality. Several
  effects, easily included in a perturbative analysis, can however
  change this smoothness. The effective theory is not only applicable
  to such situations where numerical techniques become highly involved
  but also allows one to discuss conceptual issues. For instance,
  consequences of the notoriously difficult physical inner product can
  be implemented at the effective level. This indicates that even
  physical predictions from full quantum gravity can be obtained from
  perturbative effective equations.
\end{abstract}

\maketitle

Our universe, extrapolated backwards in time, gets denser and denser,
and eventually reaches such extremes that classical general relativity
breaks down. This theoretical limitation is to be solved by a quantum
theory of gravity. Since fundamental quantum gravity implies dramatic
changes to our understanding of space and time, it can at best provide
a non-intuitive, deeply quantum description of the situation where
even concepts of time break down. More intuitive pictures can be
obtained semiclassically, where the only non-singular way a universe
can behave is as a bounce at small volume back to larger
scales. Often, models with special, mostly homogeneous, bouncing
solutions in detailed descriptions are available, but their genericity
or stability under perturbations remains uncertain. Special properties
are required for such explicit descriptions which make bounces
difficult to generalize in particular to inhomogeneities. In general,
one can then only fall back to a fully quantum
formulation. Nevertheless, under controlled perturbations bounce
scenarios could survive and present crucial insights for cosmological
scenarios. A self-consistent analysis of solutions and potential
effects of perturbations is made possible by the effective description
provided below in the framework of loop quantum gravity.

We first consider the model of a free scalar field $\phi$ in an
isotropic cosmology, which will later be seen to play an important
role for effective theory. The scale factor $a$ of the metric then
changes in proper time $\tau$ according to the Friedmann equation
$(a^{-1}\md a/\md \tau)^2=\rho$ (ignoring numerical factors) where
$\rho= \frac{1}{2}(\md\phi/\md\tau)^2$ is the energy density of
$\phi$. To quantize canonically we need to introduce canonical
variables given by the two pairs $(c,p)$ with $c=\md a/\md\tau$,
$|p|=a^2$ and $(\phi,p_{\phi})$ with $p_{\phi}=a^3\md\phi/\md\tau$.
The variable $p$ can in general take both signs unlike the scale
factor $a$ because it also contains information on the orientation of
space. Here, however, it is sufficient to assume positive $p$.
The Friedmann equation then amounts to a condition
\[
 c^2\sqrt{|p|}={\textstyle\frac{1}{2}}|p|^{-3/2}p_{\phi}^2
\]
between the phase space variables.  For a free scalar,
\begin{equation} \label{pphi}
 p_{\phi}/\sqrt{2}=\pm |cp|=:\pm H
\end{equation}
is a constant of motion, and we can view $\phi=:\sqrt{2}t$ as internal
time variable in which the dynamics of the system unfolds given by the
Hamiltonian $H$. (The value $p_{\phi}$ then plays the same role as
constant energy values in classical mechanics and the sign in
(\ref{pphi}) selects the direction of time.) A direct canonical
quantization is not fully straightforward due to the absolute value in
$H=|cp|$. Rather than using a quantization of this operator, one can
quantize just $cp$ and restrict states to the positive part of its
spectrum. Since $H$ is conserved, it suffices to pick an initial state
which is a superposition of positive eigenstates only.  Moreover, we
will be interested here in states which satisfy the semiclassicality
condition $\Delta H\ll\langle\hat{H}\rangle$ for an initial state at
large volume. Such states are sharply peaked around
$\langle\hat{H}\rangle$, which can easily be chosen positive. Thus,
for such states we can work with a quantization of $H=cp$, dropping
the absolute value.  Then, using the symmetric operator ordering
$\hat{H}=\frac{1}{2}(\hat{c}\hat{p}+\hat{p}\hat{c})$, we are lead to
the Schr\"odinger equation $\partial\psi/\partial
t=p\partial\psi/\partial p+\frac{1}{2}\psi$ for a wave function
$\psi(p,t)$.

In quantum mechanics, one usually solves first for the wave function
$\psi$ and then computes suitable expectation values relevant for
measurements. Here, it turns out to be faster and much more powerful
to solve for such expectation values directly, using a form of
generalized Heisenberg equations of motion. All information in the
system is contained in expectation values $\langle\hat{c}\rangle$,
$\langle\hat{p}\rangle$ and fluctuations or ``$n$-point functions,''
using the language of quantum field theory,
\begin{equation}
 G^{a,n}:=\langle(\hat{c}-\langle\hat{c}\rangle)^{n-a}
 (\hat{p}-\langle\hat{p}\rangle)^a\rangle_{\rm symm}
\end{equation}
for integer $n\geq 2$ and $a=0,\ldots,n$ (using again symmetric
ordering) \cite{EffAc}. Their Heisenberg-type equations of motion are
$\langle \hat{c}\rangle^{\cdot}=\langle[\hat{c},\hat{H}]\rangle/i\hbar=
\langle\hat{c}\rangle$, $\langle\hat{p}\rangle^{\cdot}=
\langle[\hat{p},\hat{H}]\rangle/i\hbar=-\langle\hat{p}\rangle$
which is easily solved by $\langle\hat{c}\rangle(t)=c_1e^{t}$,
$\langle\hat{p}\rangle(t)=c_2e^{-t}$.

To proceed similarly for the $G^{a,n}$, we need one additional
ingredient since, e.g.,
$G^{0,2}=\langle\hat{c}^2\rangle-\langle\hat{c}\rangle^2$ is not the
expectation value of an operator and the commutator
$[\langle\hat{c}\rangle^2,\hat{H}]$ is not yet defined. We can simply
do so refering to the Leibniz rule to be satisfied for commutators,
which gives $[\langle\hat{c}\rangle^2,\hat{H}]/i\hbar=
2\langle\hat{c}\rangle\langle[\hat{c},\hat{H}]\rangle/i\hbar=
2\langle\hat{c}\rangle^2$. By this procedure we have
$\dot{G}^{0,2}=\langle[\hat{c}^2-\langle\hat{c}\rangle^2, \hat{H}]
\rangle/i\hbar=2G^{0,2}$, $\dot{G}^{1,2}=0$ and
$\dot{G}^{2,2}=-2G^{2,2}$ as well as similar equations for
$\dot{G}^{a,n}$ for $n\geq 3$. Also this is easily solved by
$G^{0,2}(t)=c_3e^{-2t}$, $G^{1,2}(t)=c_4$ and $G^{2,2}(t)=c_5e^{2t}$.
Finally, fluctuations are subject to the uncertainty relation
\begin{equation} \label{uncert}
 G^{0,2}G^{2,2}-|G^{1,2}|^2\geq{\textstyle\frac{1}{4}}\hbar^2
 |\langle[\hat{c},\hat{p}]\rangle|^2
\end{equation}
which implies $c_3c_5\geq\hbar^2/4+c_4^2$.  Our solutions represent
wave packets following exactly the classical trajectory with constant
relative spread $\Delta
p/\langle\hat{p}\rangle=\sqrt{G^{2,2}}/\langle\hat{p}\rangle$.

This model is thus exactly solvable in a very strong sense: we are not
just able to obtain wave functions in closed form but the dynamics of
all $n$-point functions decouples and allows exact solutions. Such a
behavior is realized in usual quantum mechanical systems only for very
special solvable systems such as the harmonic oscillator. Its
similarity to our system is not accidental since both have quadratic,
albeit different, Hamiltonians in canonical variables. Otherwise,
non-linear coupling terms between $n$-point functions would result,
not allowing analytical solutions. Such coupling terms can easily be
seen when, e.g., adding an anharmonic potential $\hat{q}^3$ to the
harmonic oscillator. Its expectation value can be written as
$\langle\hat{p}^3\rangle=
\langle\hat{p}\rangle^3+6\langle\hat{p}\rangle G^{2,2}+6G^{3,3}$,
demonstrating the coupling between $\langle\hat{p}\rangle$ and
$G^{2,2}$. Calculating equations of motion as above shows that all of
them are now coupled non-linearly. In intuitive terms, the coupling
describes how a semiclassical wave packet spreads and deforms,
back-reacting on the peak motion given by expectation values.

A free scalar isotropic model thus serves as the analog of the
harmonic oscillator (or of a free quantum field theory) in quantum
cosmology. This observation is of crucial importance, since such free
theories always provide the basis for perturbation theory in the form
of effective equations \cite{EffAc,Karpacz}. Before discussing this
we will first introduce changes implied by a loop quantization
which we have not used so far.

The classical system and the above quantization are singular because
vanishing volume $p=0$ is approached arbitrarily closely by the
solutions. Loop quantum gravity implies modifications to the dynamics
\cite{IsoCosmo,LivRev} which lead to bounces provided that one starts
(and, in this model, then remains) in the semiclassical regime
\cite{QuantumBigBang}. This has been studied in detail recently
\cite{APS}, where the semiclassicality condition implied large values
of the Hamiltonian $H$. A loop quantization leads to a modified
Hamiltonian of the form $f(\hat{p})\widehat{\sin c}$
\cite{IsoCosmo}, with a function
\[
 f(p)\sim \left\{\begin{array}{cl} |p|(1+O(\ell_{\rm P}^2/p)) & \mbox{for
}|p|\gg\ell_{\rm P}^2\\
|p|^{-n}& \mbox{for }|p|\ll\ell_{\rm P}^2 \end{array}\right.
\]
depending on the Planck length $\ell_{\rm P}$ and a positive $n$. In
such an isotropic context, $f$ is bounded from below by a positive
number \cite{InvScale} in contrast to the classical function $p$.  If
this is used in an ``effective'' Hamiltonian $H=f(p)\sin c$, one can
easily see how bounces occur because $\sin c$ is bounded from above
such that $p$ cannot become arbitrarily small when it solves the
equation $H=f(p)\sin c$ for large constant $H$. Numerical solutions
\cite{APS} to the resulting equation for wave functions are in fact
surprisingly well described by the effective Hamiltonian $p\sin
c$: a semiclassical peak follows the effective trajectory very
precisely with negligible spread and deformations of the wave packet
in ``time'' $\phi$. This is certainly unexpected for a quantum system
where in general wave packets spread quickly and move away from the
classical trajectory.  Our {\em analytical solutions} below will
explain this behavior and {\em prove the genericity of semiclassical
  bounces beyond what is possible in a numerical analysis.}  Unlike
previous investigations, the systematic theory is well suited to a
perturbation analysis which can be performed even if strong
model-specific assumptions are relaxed. It thus provides the
technology for developing realistic bounce scenarios.

The expression $p\sin c$, using $f(p)\sim p$ for $p\gg\ell_{\rm P}^2$
which as we will see is always the case for our solutions, is not
quadratic in the variables $(c,p)$ in contrast to the classical
Hamiltonian. This seems to remove the solvability observed before, as
it generally happens when solvable Hamiltonians are modified. We are
forced to use a different Hamiltonian since the loop quantization does
not allow one to act with $c$ itself but only with the exponentials
$e^{\pm ic}$ appearing in the sine.  To deal with the new Hamiltonian,
we notice that $H$ can be reformulated in new variables $p$ and $J:=p
e^{ic}$ in such a way that it becomes even {\em linear}.  We then have
$H=-\frac{1}{2}i(J-\bar{J})$ for a Hamiltonian in non-canonical
variables $(J,p)$ with Poisson relations $\{p,J\}=-iJ$,
$\{p,\bar{J}\}=i\bar{J}$, $\{J,\bar{J}\}=2ip$. Thus, the Hamiltonian
with these new variables forms a linear algebra isomorphic to ${\rm
  sl}(2,\R)$. Classical equations of motion can easily be solved, also
taking into account the reality condition $J\bar{J}=p^2$ for our
complex variables.

If such a system remains linear after quantization, it allows a
decoupling of the infinitely many quantum variables to finitely many
coupled linear equations just as Hamiltonians quadratic in canonical
variables do. This would again make exact solutions possible.  Quite
surprisingly, the loop quantization
$\hat{H}=-\frac{1}{2}i(\hat{J}-\hat{J}^{\dagger})$ with basic
commutation relations $[\hat{p},\hat{J}]=\hbar\hat{J}$,
$[\hat{p},\hat{J}^{\dagger}]=-\hbar\hat{J}^{\dagger}$ and
$[\hat{J},\hat{J}^{\dagger}]=-2\hbar\hat{p}-\hbar^2$ does remain a
linear system. (These relations follow if the ordering
$\hat{J}=\hat{p}\widehat{\exp(ic)}$ is understood.)  The classical
algebra is simply modified by a (trivial) central extension of charge
$\hbar$. In a quantum theory, there is an additional condition
requiring that real variables are promoted to self-adjoint operators.
Such conditions determine the physical inner product used to normalize
wave functions. We do not use real variables, but the reality
condition for $J=pe^{ic}$ implies $\hat{J}\hat{J}^{\dagger}=\hat{p}^2$
to be imposed (but note $\hat{J}^{\dagger}\hat{J}\not=\hat{p}^2$ in
our ordering). Imposing the reality condition at the quantum level,
which can easily be done for quantum variables as we will see below,
corresponds to requiring that the physical inner product is used to
normalize wave functions.

Proceding as before, we derive generalized Heisenberg equations from
commutators.  This results in equations of motion
$\langle\hat{p}\rangle^{\cdot}=-\frac{1}{2}(\langle\hat{J}\rangle+
\overline{\langle\hat{J}\rangle})$ and $\langle\hat{J}\rangle^{\cdot}=
-\langle\hat{p}\rangle-\frac{1}{2}\hbar=\overline{\langle\hat{J}\rangle}^{\cdot}$
while $\langle\hat{J}\rangle-\overline{\langle\hat{J}\rangle}=2iH$ is
constant (where $H=\langle\hat{H}\rangle$). The general solution
\begin{eqnarray}
 \langle\hat{p}\rangle(t) &=& {\textstyle\frac{1}{2}}(c_1e^{-t}+c_2e^{t})-
 {\textstyle\frac{1}{2}}\hbar\\
 \langle\hat{J}\rangle(t) &=& {\textstyle\frac{1}{2}}(c_1e^{-t}-c_2e^{t})+iH
\end{eqnarray}
differs from that obtained without a loop quantization and exhibits
bouncing solutions (for $c_1c_2>0$) and singular ones ($c_1c_2<0$).
These solutions still need to be restricted by the reality condition
$\hat{J}\hat{J}^{\dagger}=\hat{p}^2$. Taking expectation values, this
tells us
\begin{equation}
 |\langle\hat{J}\rangle|^2-(\langle\hat{p}\rangle+
 {\textstyle\frac{1}{2}}\hbar)^2= {\textstyle\frac{1}{4}}\hbar^2+
 G-G^{2,2}
\end{equation}
where we introduced the dispersion variables
$G^{2,2}=\langle\hat{p}^2\rangle-\langle\hat{p}\rangle^2$ and $G=
\frac{1}{2}\langle\hat{J}\hat{J}^{\dagger}+
\hat{J}^{\dagger}\hat{J}\rangle-|J|^2=
\langle\hat{J}\hat{J}^{\dagger}\rangle- |J|^2+\hbar
p+\frac{1}{2}\hbar$, using the commutation relations. As before, one
can compute equations of motion for the dispersions and verify that
the combination $G-G^{2,2}=:\epsilon$ is constant in time. This
implies a condition for the constants in our solutions $p(t)$ and $J(t)$:
\begin{equation}
 |\langle\hat{J}\rangle|^2-(\langle\hat{p}\rangle+\hbar/2)^2= 
-4c_1c_2+ 4H^2= -\epsilon+\hbar^2/4\,.
\end{equation}
Since $\epsilon$ is a fluctuation term, it is small for semiclassical
states, in which case
$c_1c_2=H^2+\frac{1}{4}\epsilon-\frac{1}{16}\hbar^2$ is positive. In
this case, ignoring small contributions from $\epsilon$ and $\hbar$,
only bouncing solutions
\begin{equation}
\langle\hat{p}\rangle(t)=H\cosh(t-\delta)\ ,\ 
\langle\hat{J}\rangle(t)=-H(\sinh(t-\delta)+i)
\end{equation}
with $e^{\delta}:=c_1/H$ remain.  Reality conditions of the effective
theory thus prove that all solutions which are semiclassical at at
least one time are bounded away from the classical singularity at
$p=0$. (Note that this property of $\epsilon\ll H^2$ is stronger than
the condition $\Delta H\ll H$ imposed to take into account positivity
of the Hamiltonian. There are thus physical solutions with large
dispersions which reach the classical singularity, and the fact that
states which are semiclassical at one time do not reach a singularity
is non-trivial.)

\begin{figure}[ht]
\begin{center}
 \includegraphics[width=9cm,height=6cm,keepaspectratio]{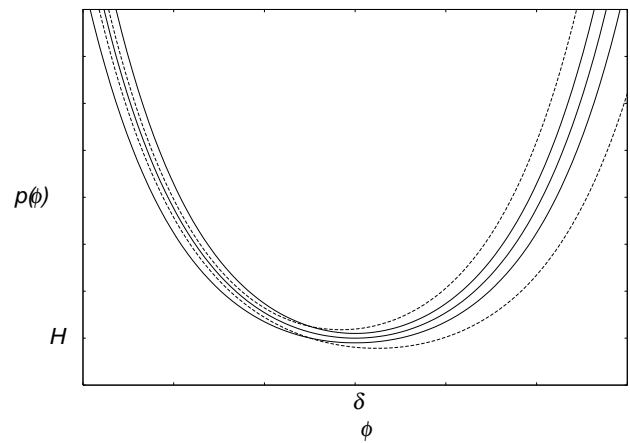}
\end{center}
\caption{Peak trajectory with surrounding area spread out by $\Delta
p$ around a bounce for $H=10\hbar$. Dashed lines show the general
behavior not assuming uniform spread. This figure is to be compared
with the numerical solution in Fig.~2 of \cite{QuantumBigBang}.}
\label{EffBounce}
\end{figure}

Similarly, we can explicitly determine the spread parameters. We use
the second order variables
$G^{0,2}=\langle\hat{J}^2\rangle-\langle\hat{J}\rangle^2$,
$G^{1,2}=\frac{1}{2}(\langle\hat{p}\hat{J}\rangle+
\langle\hat{J}\hat{p}\rangle)-
\langle\hat{p}\rangle\langle\hat{J}\rangle$ and
$G^{2,2}=\langle\hat{p}^2\rangle-\langle\hat{p}\rangle^2$.  (The
operator $\hat{J}^{\dagger}$, although it appears in the Hamiltonian,
is not included in quantum variables here since we can use the
commutation relations and the reality condition to express any
expectation value containing $\hat{J}^{\dagger}$ through quantum
variables of the same or lower order not containing
$\hat{J}^{\dagger}$.)  By the general scheme we have equations of
motion $\dot{G}^{0,2}=-2G^{1,2}$, $\dot{G}^{1,2}=
-\frac{1}{2}G^{0,2}-\frac{3}{2}G^{2,2}- \frac{1}{8}\hbar^2$,
$\dot{G}^{2,2}=-2G^{1,2}$. Thus, $G^{0,2}-G^{2,2}=2c_3$ is constant
while
\begin{eqnarray*}
 G^{2,2} &=& {\textstyle\frac{1}{2}}(c_4e^{-2t}+c_5e^{2t})-
 {\textstyle\frac{1}{2}}c_3-
{\textstyle\frac{1}{16}}\hbar^2\\
 G^{1,2} &=& {\textstyle\frac{1}{2}}(c_4e^{-2t}-c_5e^{2t})\,.
\end{eqnarray*}
%
The constants $c_i$ are restricted again by the
uncertainty relation which now, in non-canonical variables, reads
\begin{equation}
 G^{0,2}G^{2,2}-|G^{1,2}|^2\geq{\textstyle\frac{1}{4}}\hbar^2|J|^2\,.
\end{equation}
For $t\to\pm\infty$, this is easy to evaluate and satisfied only if
$c_3$, $c_4$ and $c_5$ are positive. Moreover, {\em assuming
$H\gg\hbar$ and near saturation before as well as after the bounce} one has
$c_4e^{-2\delta}\approx\hbar H\approx c_5e^{2\delta}$. With this,
\begin{eqnarray}
 (\Delta p(t))^2=G^{2,2}(t)\approx \hbar H\cosh(2(t-\delta))
\end{eqnarray}
and not only the peak trajectory but also the spread of
the wave packet is symmetric around $t=\delta$. But only under
the two stated conditions does the solution automatically become as
coherent after the bounce as it was before (Fig.~\ref{EffBounce}).

Our solutions in the given ordering are exact \footnote{We had to drop
  the absolute value of $\hat{H}$ to derive those solutions, which is
  well justified for the states we consider here. If there were
  coupling terms between expectation values and dispersions, on the
  other hand, the semiclassical dynamics would change considerably.}
quantum solutions where $\langle\hat{p}\rangle(t)$ is unaffected by
spreading. This shows clearly why the naive effective formulation,
obtained by simply replacing $c^2$ by $\sin^2c$ in the classical
Hamiltonian, agrees so well with numerical simulations of a related
free scalar model \cite{APS}. But the smoothness and symmetry of the
bounce requires additional assumptions which brings us to its
robustness. As mentioned, the model is fully analogous to the harmonic
oscillator as far as the decoupling of all its quantum variables is
concerned. Using exclusively one solvable model to draw conclusions
for the quantum nature of the big bang can then be as misleading as
using the harmonic oscillator to determine the general behavior in
quantum mechanics would be. Under any new ingredient to make the model
more realistic coupling terms between all variables will arise. States
then do spread and deform with complicated non-linear back-reaction on
the expectation values.  We thus need to discuss the expected more
general behavior, before coming to effective theory as a means to
study stability and develop realistic scenarios.

There are several examples for generalizations.  First, while
$H\gg\hbar$ is necessary for a semiclassical state at large volume of
a homogeneous universe with large matter content, several effects in
{\em inhomogeneous} models remove this condition. Then, $H$ is a
spatial function and its local values relevant for loop variables can
be small even with a large matter content. (See
\cite{InhomLattice,QuantCorrPert} for properties of inhomogeneous
states and operators defined in a regime around a homogeneous
reference metric.)  Moreover, with many more degrees of freedom,
inhomogeneous situations allow the emergence of a semiclassical state
at large volume, e.g.\ through decoherence, even though the state
around the bounce can be highly non-semiclassical.  This will
complicate an analysis based only on the condition that the universe
be semiclassical at one late time. 

But already in homogeneous models, several new effects can arise if
$H$ is not large, as in those inhomogeneous cases.  The bounce scale
will be smaller and quantum corrections to the $p$-dependence of the
constraint through the function $f(p)$ will appear. Since these
corrections are higher powers in $p$, coupling terms between $p$ and
all quantum variables arise. At some point, due to the boundedness of
$f(p)$, no bounce occurs at all when $H$ is small enough. (A similar
removal of the bounce due to structural modifications to the
equations, loosely related to potential consequences of inhomogeneous
states on small scales, has recently been explored in \cite{Shallow}.)
Moreover, the behavior becomes more sensitive to the factor ordering
of the constraint.  Ordering all $\hat{p}$ to the right, for instance,
leads to an additional contribution $\hbar\hat{p}^{-1}\hat{J}$ to the
constraint.  This is of the order $\hbar$ and would not change the
above analysis much when $H$ is large, but it does become relevant for
smaller $H$.  It implies coupling terms of the form $-\hbar
G^{1,2}/\langle\hat{p}\rangle^2$.  In addition, there are corrections
if model specifics are changed.  Adding a cosmological constant or
positive curvature adds coupling terms to the Hamiltonian, but near
the bounce they are not dominant. A potential for the scalar field,
however, is more crucial because it leads to a time-dependent
Hamiltonian.

In particular the latter change can have strong implications depending
on the form of the potential (see, e.g., the last example in
\cite{Time}). The most important ingredient to introduce is
inhomogeneities and their evolution, at least as perturbations around
a large isotropic background, which is necessary both for a stability
analysis and for cosmological predictions. Inhomogeneities add not
only new degrees of freedom but also additional coupling terms between
the variables.  All this is difficult to analyze numerically, and
generalized Heisenberg equations as used here can also become
complicated to solve exactly. But all these changes are
straightforwardly implemented in a general perturbation scheme around
the solvable free scalar model identified here. The model behavior is
valid when only large semiclassical scales are involved and
interactions are negligible, but it receives corrections when smaller
scales are reached or interactions introduced.  The resulting picture,
accessible perturbatively, can be quite different from the smooth
bouncing solutions obtained for large $H$ and isotropic states.

This is not surprising because inhomogeneous situations provide many
degrees of freedom over which excitations can be distributed which
happens in particular close to a classical singularity. The generic
behavior is then different from an isotropic bounce, an observation
which has also been made in the context of string theory
\cite{StringInHom}. General singularity resolution can be achieved by
mechanisms taking into account deep quantum behavior which does not
necessarily provide intuitive pictures
\cite{Sing,HomCosmo,Spin,SphSymmSing}.  While such a direct quantum
analysis is difficult in explicit terms, effective theory does make it
feasible by perturbing around an exact model. Given a Hamiltonian
$H=H_0+ H_1(p,J;t)$ with the solvable one $H_0$ and possibly time
(i.e., scalar field) dependent coupling terms $H_1$, we find
perturbative solutions $\langle\hat{p}\rangle(t)=p_0(t)+p_1(t)$ where
$p_0$ is a free solution as above and $p_1$ is obtained by integrating
\begin{equation}
 \dot{p}_1=
 -i\hbar^{-1}\langle[\hat{p},\hat{H}_1]\rangle|_{\langle\hat{p}\rangle=p_0(t),
\langle\hat{J}\rangle=J_0(t),G^{a,n}=G_0^{a,n}(t)}\,.
\end{equation}
The right hand side is known in terms of zero order
solutions. This procedure agrees with perturbation expansions in the
interaction picture of quantum field theory.  Indeed, effective
equations derived by perturbation theory around our solvable model are
in complete analogy with low energy effective actions in particle
physics \cite{EffAc,Karpacz}.  

We thus provide the first {\em systematic scheme for a well-defined
  evolution of perturbations through bounces}. It allows us to
approach the strong quantum regime and to determine if a
semiclassical description would break down.  One can, for instance, start
with large $H$ and then decrease its value to see corresponding
changes in the wave function. (Although different in its nature, this
parameter thus plays a role similar to others which have already been
used in loop quantum cosmology \cite{Ambig,Inflation}.  While the
parameter should not be large for realistic inhomogeneous bounces, it
provides a technical tool to separate some quantum effects from
others.) In perturbation theory one can then deal with quantum
corrections arising from several coupling terms which may be present.

Effective theory is thus a powerful tool where direct numerical
simulations become more involved. One can obtain directly quantities
of interest such as expectation values and fluctuations without taking
the ``detour'' of wave functions. What is particularly interesting for
full quantum gravity is that even conceptual issues can be addressed.
The problem of time does not arise because one is perturbing around a
model which is explicitly parameterized. Similarly, observables can be
evaluated perturbatively. Even the notoriously difficult issue of the
physical inner product, i.e.\ the issue of how to select normalizable
wave functions, can be addressed at the effective level as already
seen: we used reality conditions
to select the physically viable effective solution which turned out to
be bouncing. This corresponds to the correct normalizable wave
function \cite{APS}. This solution is uniquely selected by referring
only to quantum variables side-stepping explicit states, an
observation which is encouraging for effective developments in full
quantum gravity. In effective equations, one can then implement the
physical inner product order by order on effective solutions. In
particular, this argument applied to general dynamical expressions
provides the strongest indication so far that repulsive effects which
have been observed throughout loop cosmological models \cite{WS:MB}
are active even in a full setting.  Although they are not strong
enough to lead to general bounces at smaller scales of generic states,
we have shown that perturbative regimes of inhomogeneous bounces do
exist. Whether or not this is sufficient for our universe can be
tested in detailed analyses which are now being undertaken. Through
effective theory, all issues relevant for physical predictions can
thus be addressed.

{\bf Acknowledgements:} The author thanks A.\ Ashtekar, T.\ Pawlowski
and P.\ Singh for discussions.  This work was supported by NSF
grant PHY0554771.


\begin{thebibliography}{10}

\bibitem{EffAc}
M. Bojowald and A. Skirzewski, Rev.\ Math.\ Phys.\ {\bf 18}, 713
(2006); A.\ Skirzewski, Ph.D.\ Thesis (HU Berlin, 2006).

\bibitem{Karpacz}
M. Bojowald and A. Skirzewski, Int.\ J.\ Geom.\ Meth.\ Mod.\ Phys.\
{\bf 4}, 25 (2007), hep-th/0606232.

\bibitem{IsoCosmo}
M. Bojowald, Class.\ Quantum Grav. {\bf 19},  2717  (2002).

\bibitem{LivRev}
M. Bojowald, Living Rev.\ Relativity {\bf 8},  11  (2005).

\bibitem{QuantumBigBang}
A. Ashtekar, T. Pawlowski, and P. Singh, Phys.\ Rev.\ Lett. {\bf 96},  141301
  (2006).

\bibitem{APS}
A. Ashtekar, T. Pawlowski, and P. Singh, Phys.\ Rev.\ D {\bf 73},
124038 (2006).

\bibitem{InvScale}
M. Bojowald, Phys.\ Rev.\ D {\bf 64},  084018  (2001).

\bibitem{InhomLattice}
M. Bojowald, Gen.\ Rel.\ Grav. {\bf 38},  1771  (2006).

\bibitem{QuantCorrPert}
M. Bojowald, H. Hern\'andez, M. Kagan, and A. Skirzewski, Phys.\ Rev.\
D {\bf 74}, 123512 (2006).

\bibitem{Shallow}
P.\ Laguna, Phys.\ Rev.\ D {\bf 75}, 024033 (2007).

\bibitem{Time}
M. Bojowald, P. Singh, and A. Skirzewski, Phys.\ Rev.\ D {\bf 70},  124022
  (2004).

\bibitem{StringInHom}
T. Hertog and G.~T. Horowitz, JHEP {\bf 0504},  005  (2005).

\bibitem{Sing}
M. Bojowald, Phys.\ Rev.\ Lett. {\bf 86},  5227  (2001).

\bibitem{HomCosmo}
M. Bojowald, Class.\ Quantum Grav. {\bf 20},  2595  (2003).

\bibitem{Spin}
M. Bojowald, G. Date, and K. Vandersloot, Class.\ Quantum Grav. {\bf 21},  1253
   (2004).

\bibitem{SphSymmSing}
M. Bojowald, Phys.\ Rev.\ Lett. {\bf 95},  061301  (2005).

\bibitem{Ambig}
M. Bojowald, Class.\ Quantum Grav. {\bf 19},  5113  (2002).

\bibitem{Inflation}
M. Bojowald, Phys.\ Rev.\ Lett. {\bf 89},  261301  (2002).

\bibitem{WS:MB}
M. Bojowald,  in {\em 100 Years of Relativity -- Space-Time Structure: Einstein
  and Beyond}, edited by A. Ashtekar (World Scientific, Singapore, 2005), pp.\
  382--414.

\end{thebibliography}
\end{document}